\documentclass[12pt]{article}
\pdfoutput=1
\usepackage{textcomp}
\usepackage{color}
\usepackage{putex}
\usepackage{autobreak}
\usepackage{indentfirst}
\usepackage{graphicx}
\usepackage{float}
\graphicspath{{plots/}}
\usepackage{tabularx}
\usepackage{caption}
\usepackage{tikz, pgfplots}
\usetikzlibrary{arrows.meta, calc, positioning}
\usetikzlibrary{decorations.pathmorphing}
\usepackage{amsmath}
\numberwithin{equation}{section}
\usepackage{mathtools}
\usepackage{cancel}
\usepackage{array}
\usepackage{subcaption}
\usepackage{epstopdf}
\usepackage{enumerate}
\usepackage{cite}
\usepackage{youngtab}
\usepackage{tensor}
\usepackage{slashed}
\usepackage[aligntableaux=center]{ytableau}
\usepackage{CJKutf8}
\usepackage[utf8]{inputenc}
\usepackage{rotating}
\usepackage{multirow}
\usepackage[
      colorlinks=true,
      linkcolor=red,
      urlcolor=red,
      filecolor=black,
      citecolor=red,
      ]{hyperref}
\usepackage[noabbrev]{cleveref}
\begin{document}
\institution{forbiddencity}{Kavli Institute for Theoretical Sciences, University of Chinese Academy of Sciences, \cr Beijing 100190, China.}
\title{
Spin-$2$ operators in AdS$_2$/CFT$_1$
}
\authors{
Konstantinos C.~Rigatos\footnote{{\hypersetup{urlcolor=black}\href{mailto:92konstantinos10@gmail.com}{92konstantinos10@gmail.com}}}
}
\abstract{
We study spin-2 fluctuations around an infinite family of warped backgrounds of the form $\text{AdS}_2 \times \text{S}^2 \times \text{T}^4 \times \text{S}^1 \times \mathcal{I}_{\rho}$ in the type IIB theory, which possesses $\mathcal{N}=4$ supersymmetry. We find that there exists a special set of solutions which is independent of the background data; the minimal universal class. This particular class of solutions corresponds to operators with scaling dimension $\Delta = \ell + 1$, with $\ell$ being the quantum number of the angular-momentum on the $\text{S}^2$. Using the mode solutions of these spin-2 states, we compute the central charge of the dual field theories. We comment on the relation of our results to the seed $\text{AdS}_3 \times \text{S}^2 \times \text{T}^4 \times \mathcal{I}_{\rho}$ backgrounds. 
}
\date{\today}
\maketitle
{
\hypersetup{linkcolor=black}
\tableofcontents
}
\newpage
\section{Prolegomena}
The AdS/CFT duality \cite{Maldacena:1997re,Witten:1998qj,Gubser:1998bc} represents a primo realization of the holographic principle. The correspondence postulates that specific string theories with an AdS factor in their geometry are equivalent descriptions of field theories on the boundary of the space. Since its  discovery, there's been a renewed interest in the systematic study of superconformal field theories across different dimensions and with various amounts of supersymmetry and the construction of the dual AdS descriptions. 

The space of consistent and non-trivial half-maximal superconformal field theories is very rich in structures, in addition to being well under control due to the high amount of symmetry. It has received a lot of attention with various studies having been performed using different approaches and techniques. 

For the four-dimensional $\mathcal{N}=2$ theories first suggested by Gaiotto \cite{Gaiotto:2009we}, the duals were derived in \cite{Lin:2004nb,Gaiotto:2009gz}, with further studies in this context to have been pursued in \cite{Reid-Edwards:2010vpm,Aharony:2012tz,Lozano:2016kum,Nunez:2018qcj,Nunez:2019gbg}. The case of five-dimensional field theories with $\mathcal{N}=1$ supersymmetry has been thoroughly studied both from the field theory and the supergavity points of view \cite{Seiberg:1996bd,Morrison:1996xf,Ferrara:1998gv,Brandhuber:1999np,Lozano:2012au,Bergman:2012kr,DHoker:2016ysh,Lozano:2013oma,DHoker:2016ujz,DHoker:2017mds,DHoker:2017zwj,Fluder:2018chf,Uhlemann:2019ypp,Uhlemann:2020bek,Legramandi:2021uds,Akhond:2022awd,Akhond:2022oaf,Penin:2019jlf}. Progress has, also, been made in the context of six-dimensional $\mathcal{N}=(1,0)$ theories \cite{Brunner:1997gf,Hanany:1997gh}, which have known supergravity duals in the context of massive type IIA \cite{Apruzzi:2015wna,Apruzzi:2013yva,Gaiotto:2014lca,Cremonesi:2015bld,Nunez:2018ags,Filippas:2019puw,Bergman:2020bvi}. 

There has been intense reasearch and progress not only for the case of four and higher dimensional field theories, but also for lower dimensional ones. Three-dimensional $\mathcal{N}=4$ theories have been studied \cite{Akhond:2022oaf,Gaiotto:2008ak,Bachas:2019jaa,Lozano:2016wrs,DHoker:2008lup,DHoker:2007hhe,Assel:2011xz,Akhond:2021ffz} and the very appealing case of two-dimensional superconformal field theories and their $\text{AdS}_3$ dual backgrounds has gained its own attention and led to new, interesting and exciting results \cite{Lozano:2015bra,Macpherson:2018mif,Legramandi:2019xqd,Couzens:2017way,Couzens:2019wls,Couzens:2020aat,Lozano:2019emq,Lozano:2019zvg,Lozano:2019jza,Lozano:2019ywa,Lozano:2020bxo,Filippas:2020qku,Rigatos:2020igd,Filippas:2019ihy}.

In the above beautiful picture, there is an enigmatic corner. The case being made for the $\text{AdS}_2/\text{CFT}_1$ duality which is the less understood example in the context of holography due to some subtleties that arise in that case and are not present in the higher-dimensional holographic duals \cite{Maldacena:1998uz,Almheiri:2014cka,Denef:2007yt,Harlow:2018tqv}. When considering the effect of backreaction for excitations of finite-energy, one obtains a divergence at one of the two asymptotic boundaries. Moreover, since global $\text{AdS}_2$ possesses two disconnected boundaries, it seems that the putative holographic description should be given in terms of two copies of a one-dimensional theory. This is contradiction to the calculations of black hole entropy in the context of string theory.

Be that as it may, owing to the similarities between the $d=1$ and $d=2$ superconformal algebras, or the geometric connections between $\text{AdS}_2$ and $\text{AdS}_3$ spaces, there is, by now, some progress towards the better understanding of the picture \cite{Cvetic:2000cj,Gauntlett:2006ns,DHoker:2007mci,Gupta:2008ki,Chiodaroli:2009yw,Chiodaroli:2009xh,Kim:2013xza,Corbino:2017tfl,Dibitetto:2018gbk,Dibitetto:2018gtk,Corbino:2018fwb,Hong:2019wyi,Dibitetto:2019nyz,Corbino:2020lzq}. Studies pertaining to $\text{AdS}_2$ geometries that have been motivated either black holes physics, or a purely geometric/field theoretic viewpoint, or both \cite{Strominger:1998yg,Cadoni:1999ja,Balasubramanian:2009bg,Azeyanagi:2007bj,Castro:2014ima,Cvetic:2016eiv,Anninos:2017hhn,Anninos:2019oka,Mirfendereski:2020rrk,Hartman:2008dq,Alishahiha:2008tv,Bena:2018bbd,Heidmann:2022zyd}.

An interesting antithesis was presented in \cite{Lunin:2015hma} where a broad class of geometries that are regular was constructed. This class asymptotes to $\text{AdS}_2 \times \text{S}^2$. In these wormhole-like solutions, the presence of two boundaries was a feature rather than a bug of the construction. The expansion of these solutions around the $\text{AdS}_2 \times \text{S}^2$ background results to perturbations similar in spirit to the ones we will consider in this work, although in a different string theory embedding. 

Recently the classification of the $\text{AdS}_2$ string backgrounds and the associated dual field theory picture got enlarged \cite{Lozano:2020txg,Lozano:2020sae,Lozano:2021rmk,Lozano:2021fkk,Lozano:2022vsv}. More specifically, the authors in \cite{Lozano:2020txg} derived $\text{AdS}_2$ string geometries that possess one $\text{SU}(2)$ isometry. This leads to the study of $\mathcal{N}=4$ superconformal quantum mechanics. The background solutions, which are constructed in the type IIB theory, are trustworthy as dual descriptions of the $d=1$ conformal field theories, so long as the number of nodes of the quiver and the ranks of each gauge group are large.

Since $\text{AdS}_3$ and $\text{AdS}_2$ share structural similarities, this suggests that some of the holographic studies in the $\text{AdS}_3$ backgrounds can naturally be extended to the case of $\text{AdS}_2$ solutions. It also suggests, that we should be able to use $\text{AdS}_3/\text{CFT}_2$ results, in order to obtain information on $\text{AdS}_2/\text{CFT}_1$ and of course vice versa. 

Related to a better understanding of any superconformal field theory, a very important aspect in the study of the spectrum of operators, and the understanding of how they organise themselves into the different representations of the underlying algebra. 

By virtue of the AdS/CFT, in order to describe the spectrum of the gauge-invariant field theory operators we need to consider the linearised fluctuations of the bulk fields around a given supergravity solution. Computing all the linearised fluctuations for the full supergravity background is a daunting task, see the works of \cite{Kim:1985ez,Deger:1998nm} for the computations of the full Kaluza-Klein spectrum in $\text{AdS}_5 \times \text{S}^5$ and $\text{AdS}_3 \times \text{S}^3$ respectively.

Be that as it may, the work of \cite{Bachas:2011xa} proved that if we consider only the spectrum of spin-2 operators and the situation simplifies dramatically. In essence the authors showed that spin-2 fluctuations, which are perturbations of the geometry, satisfy an equation that depends only on data that we can extract by examining the underlying geometry of the background supergravity configuration. While that paper was in the context of an $\text{AdS}_4$ solution, generalizing the logic is quite straightforward. This paved the way for many new interesting results on the spectrum of holographic spin-2 operators in the contexts of $\text{AdS}_7/\text{CFT}_6$ \cite{Passias:2016fkm}, $\text{AdS}_6/\text{CFT}_5$ \cite{Passias:2018swc,Gutperle:2018wuk}, $\text{AdS}_5/\text{CFT}_4$ \cite{Chen:2019ydk,Itsios:2019yzp}, $\text{AdS}_4/\text{CFT}_3$ \cite{Richard:2014qsa,Klebanov:2009kp} and $\text{AdS}_3/\text{CFT}_2$ \cite{Speziali:2019uzn,Lima:2022hji}. 

Taking that path, we will be extending the catalogue of the previous spin-2 results by studying some of the linearised fluctuations that arise in the $\text{AdS}_2 \times \text{S}^2 \times \text{CY}_2 \times \mathcal{I}_{\psi} \times \mathcal{I}_{\rho}$ supergravity background recently constructed in \cite{Lozano:2020txg}. The dual picture is described by $\mathcal{N}=4$ superconformal quantum mechanics. Our motivation is to holographically construct part of the spectrum of operators of those theories and to examine how much information we can obtain for the seed $\text{AdS}_3$ from studying the $\text{AdS}_2$ solutions. As byproducts of our studies, we derive a bound on the mass of the modes and the central charge of the associated theories.

The structure of this work is the following: in \cref{sec: sugra} we review and discuss the basic aspects of the type IIB $\text{AdS}_2$ solutions that were derived in \cite{Lozano:2020txg} as well as some aspects of the dual field theory interpretation. Then we proceed to \cref{sec: spin_2} to derive the equation that governs the dynamics of the spin-2 fluctuations of the background metric. These are the transverse-traceless perturbations along the non-compact part of the geometry under consideration and they correspond to massive rank-2 tensors. In \cref{sec: universal} we derive a bound for the mass and we associate it to the unitarity bound of conformal dimension of the associated operators. We are, also, able to obtain a special class of solutions that saturates the bound and is independent of functions tha define the background, hence this class of solutions is present in all quivers related to these type IIB solutions. The modes derived by this special class of solutions are dubbed minimal universal class. Moving along, in \cref{sec: central_charge} we use the action of spin-2 fluctuations to compute the holographic central charge of the associated theories, and we find agreement with an independent computation. \Cref{sec: multiplets} contains discussion on the field theory implications of our studies. In \cref{sec: non_uni} we provide an example of a non-universal solution, while \cref{sec: relation_ads3} contains discuss on the relation of our results to preciously obtained studies in the context of the $\tfrac{1}{4}$-BPS $\text{AdS}_3$ seed solutions. We summarize our findings and conclude in \cref{sec: epilogue}. 
\section{The supergravity backgrounds and the dual field theories} \label{sec: sugra}
We would like to begin by reviewing the basic characteristics of the geometries derived in \cite{Lozano:2020txg} and their associated dual field theories. These are solutions in the type IIB theory with the form $\text{AdS}_2 \times \text{S}^2 \times \text{CY}_2 \times \mathcal{I}_{\psi} \times \mathcal{I}_{\rho}$, where $\mathcal{I}_{\rho}$ is a finite interval. They were originally derived by considering a T-duality transformation along an $AdS$ direction of the massive type IIA solutions that have the form $\text{AdS}_3 \times \text{S}^2 \times \text{CY}_2 \times \mathcal{I}_{\rho}$ originally constructed in \cite{Lozano:2019emq}. 
	\subsubsection*{The supergravity description}
The AdS$_2$ solution derived in \cite{Lozano:2020txg} is described by the geometry
\begin{equation}\label{eq: ads2_geo}
ds^2 = \frac{u}{\sqrt{\hat{h}_4 h_8}} \left( \frac{1}{4} ds^2_{\text{AdS}_2} + \frac{\hat{h}_4 h_8}{4 \hat{h}_4 h_8 + (u')^2} ds^2_{\text{S}^2} \right) + \sqrt{\frac{\hat{h}_4}{h_8}} ds^2_{\text{CY}_2} + \frac{\sqrt{\hat{h}_4 h_8}}{u} (d \rho^2 +  d \psi^2 )  \, ,
\end{equation}
and is equipped with the following fields in the NS-NS sector 
\begin{equation}\label{eq: ads2_nsns}
\begin{split}
e^{- 2 \Phi}	&= \frac{h_8}{4\hat{h}_4} \left(4 \hat{h}_4 h_8 + (u')^2 \right) \, , \\ 
H_{3}			&= \frac{1}{2} d \left( - \rho + \frac{u u'}{4 \hat{h}_4 h_8 + (u')^2} \right) \wedge \vol_{S^2} + \frac{1}{2}\vol_{\text{AdS}_2} \wedge d \psi \, ,
\end{split}
\end{equation}
with $\psi$ being the T-dual coordinate that takes values in the range $[0,2\pi]$. Here $\Phi$ is the dilaton, $H_3$ the NSNS three-form and the metric is written in the string frame.  The functions $u$, $\hat{h}_4$, $h_8$ are functions only of the $\rho$ coordinate, and we have used a prime to denote a derivative with respect to $\rho$-coordinate.

A more complicated variant of the above description is when $\hat{h}_4$ has an explicit dependence on $(\rho, \text{CY}_2)$. This more general form of backgrounds is briefly described \cite[appendix A]{Lozano:2020txg}. For us, it is sufficient to say that the case $\hat{h}_4=\hat{h}_4(\rho, \text{CY}_2)$ is associated to a function in the seed $\text{AdS}_3$ backgrounds, which was named $H_2$. Upon considering that $H_2$ vanishes, we obtain the solutions shown above, which will be the case of interest for us here.

The RR sector our solutions reads
\begin{equation}\label{eq: ads2_rr}
\begin{split}
F_{1} 	&= h_8' d \psi \, , \\ 
F_{3}	&=  - \frac{1}{2} \left( h_8 - \frac{h_8' u' u}{4 h_8 \hat{h}_4+(u')^2} \right) \vol_{\text{S}^2} \wedge d \psi + \frac{1}{4} \left( d \left( \frac{u'u}{2 \hat{h}_4} \right) + 2 h_8 d \rho \right) \wedge \vol_{\text{AdS}_2} \, , \\
F_{5}	&= -(1 + \star) \, \hat{h}_4' \, \vol_{\text{CY}_2} \wedge d \psi \, ,\\ 
		&= - \hat{h}_4' \, \vol_{\text{CY}_2} \wedge d \psi + \frac{\hat{h}_4' h_8 u^2}{4 \hat{h}_4 (4 \hat{h}_4 h_8 + (u')^2)} \vol_{\text{AdS}_2} \wedge \vol_{\text{S}^2} \wedge d \rho \, , \\
\end{split}
\end{equation}
and the higher-form RR fields are given by $F_{7} = - \star F_{3}$ and $F_{9} = \star F_{1}$. 

The Type IIB equations of motion are satisfied by imposing the BPS equations and Bianchi identities: 

\begin{equation}
u''=0
\end{equation}

and 

\begin{equation}\label{eq: Bianchi_no_sources}
\hat{h}_4''= 0 \,	,	\qquad h_8''=0	\,	.
\end{equation}

Hence, it is easy to see that the three functions $u$, $\hat{h}_4$ and $h_8$ that are the warp factors and fully determine the $AdS_2$ solutions are at most linear in the $\rho$-coordinate. This holds true away from explicit brane sources. In the presence of branes, there is a violation of the Bianchi identities as the right-hand sides of \cref{eq: Bianchi_no_sources} receive infinite contributions in the form of delta functions. This leads to $\hat{h}_4$ and $h_8$ being piecewise linear functions. 

The above background can be depicted pictorially by the brane intersection presented in \cref{tab: brane_scan}

\begin{table}[H]
	\begin{center}
		\begin{tabular}{|c|c|c|c|c|c|c|c|c|c|c|c|}
			\hline	
 &&&&&&&&&&\\[-0.95em] 	    
			&	$x^0$	& $x^1$ 		& $x^2$ 		& $x^3$			& $x^4$ 		& $x^5$ 		& $x^6$ 		& 	$x^7$ 		& 	$x^8$ 		& 	$x^9$ 			\\ \hline \hline
	D1 		& 	--- 	& $\bullet$		& $\bullet$		& $\bullet$ 	& $\bullet$ 	& --- 			& $\bullet$  	&   $\bullet$	& 	$\bullet$  	&  $\bullet$ 		\\ \hline
	D3 		& 	---		& $\bullet$		& $\bullet$ 	& $\bullet$ 	& $\bullet$ 	& $\bullet$		& 	--- 		& 	--- 		& 	--- 		&  $\bullet$ 		\\ \hline
	D5 		& 	---		& --- 			& 	--- 		& --- 			& --- 			& --- 			& $\bullet$  	&   $\bullet$	&   $\bullet$	&  $\bullet$ 		\\ \hline
	D7 		& 	---		& ---			&	---  		& --- 			& --- 			& $\bullet$		& ---  			& 	--- 		& 	--- 		&  $\bullet$ 		\\ \hline
	NS5 	& 	---		& ---			&	---  		& --- 			& --- 			& $\bullet$		& $\bullet$  	&   $\bullet$	&   $\bullet$	& 	---  			\\ \hline
	F1 		&	---		& $\bullet$		& $\bullet$		& $\bullet$		& $\bullet$		& $\bullet$		& $\bullet$		& 	$\bullet$	& 	$\bullet$	& 	--- 			\\ \hline
		\end{tabular} 
\caption{A pictorial depiction of the brane-scan that describes the backgrounds given by \cref{eq: ads2_geo,eq: ads2_nsns,eq: ads2_rr}. In our notation (---) stands for the brane extending along that particular direction, while $\bullet$ denotes a coordinate is transverse to the brane. In the above, $x^0$ is the time direction of the ten-dimensional spacetime. The directions $(x^1, \dots , x^4)$ span the Calabi-Yau two-fold, $x^5$ is the direction associated with $\rho$, $(x^6, x^7, x^8)$ are the transverse directions realising the $\text{SO}(3)$-symmetry of the $\text{S}^2$, and $x^9$ is the $\psi$ direction.}
\label{tab: brane_scan}
\end{center}
\end{table}

Following \cite{Lozano:2020txg}, we will be interested in the case of a finite interval $\mathcal{I}_{\rho}$, where both the $\hat{h}_4$ and $h_8$ vanish at the endpoints of the interval. Therefore, to start fixing conventions, let us call $\mathcal{I}_{\rho} = [0, \rho^*]$ and $\hat{h}_4(\bar{\rho}) = h_8(\bar{\rho}) = 0$, when $\bar{\rho}$ is equal to $0$ or  $\rho^*$. It is convenient \cite{Lozano:2020txg} to set $\rho^* = 2 \pi (P + 1)$, with $P$ being a large integer. The remaining function, $u$, vanishes only at $\rho = 0$\footnote{there exists, also, the possibility that $u = \texttt{constant}$ everywhere, which we will not discuss in this work.}. The general form for $\hat{h}_4$, $h_8$ and $u$ is given by the following expressions:
  \begin{equation}\label{eq: def_h4}
    \hat{h}_4(\rho) =
  \Upsilon  \begin{cases*}
      \frac{\beta_0}{2 \pi} \rho & $ 0 \leq \rho \leq 2 \pi$  \\
      \beta_0 + \dots + \beta_{k-1}  + \frac{\beta_k}{2 \pi}(\rho - 2 \pi k)   & $2 \pi k < \rho \leq 2 \pi (k+1) \, , \quad k = 1, \cdots, P-1$ \\
      \alpha_P - \frac{\alpha_P}{2 \pi} (\rho - 2 \pi P) & $2 \pi P < \rho \leq 2 \pi (P+1) \, , $
    \end{cases*}
\end{equation}
 \begin{equation}\label{eq: def_h8}
    h_8(\rho) =
    \begin{cases*}
      \frac{\nu_0}{2 \pi} \rho & $ 0 \leq \rho \leq 2 \pi$  \\
      \nu_0  + \dots + \nu_{k-1} + \frac{\nu_k}{2 \pi}(\rho - 2 \pi k)   & $2 \pi k < \rho \leq 2 \pi (k+1) \, , \quad k = 1, \cdots, P-1$ \\
      \mu_P - \frac{\mu_P}{2 \pi} (\rho - 2 \pi P) & $2 \pi P < \rho \leq 2 \pi (P+1) \, . $
    \end{cases*}
\end{equation}

In the above, the quantities $(\alpha_k,\beta_k, \mu_k, \nu_k)$ are integration constants, and continuity of $\hat{h}_4$ and $h_8$ determines that:
\begin{equation}\label{eq: defs_alpha_beta}
\alpha_k=\sum_{j=0}^{k-1} \beta_j \,	,	\qquad \mu_k= \sum_{j=0}^{k-1}\nu_j		\,		.
\end{equation} 

Finally, the function $u(\rho)$ is given by:
\begin{equation}\label{eq: def_u}
u = \frac{b_0}{2 \pi} \rho \, .
\end{equation}

By performing the asymptotic expansions at the endpoints of the $\mathcal{I}_{\rho}$ we find that as $\rho=0$ the background is regular. At the other endpoint, denoted by $\rho=2\pi(P+1)-x$, we perform the expansion for small $x$, and we obtain:

\begin{equation}\label{eq: asymptotes}
ds^2\sim \frac{1}{x} ds^2_{\text{AdS}_2} + x(dx^2+  d\psi^2 + ds^2_{\text{S}^2} ) + ds^2_{\text{CY}_2}	\,	,	\qquad 	e^{-2\Phi} \sim 1		\,	.	
\end{equation}

The above describes the superposition of O1- and O5-planes that extend along $\text{AdS}_2$ and are smeared on the $\text{CY}_2$ and $\text{AdS}_2 \times \text{CY}_2$ respectively.
	\subsubsection*{The dual field theories}
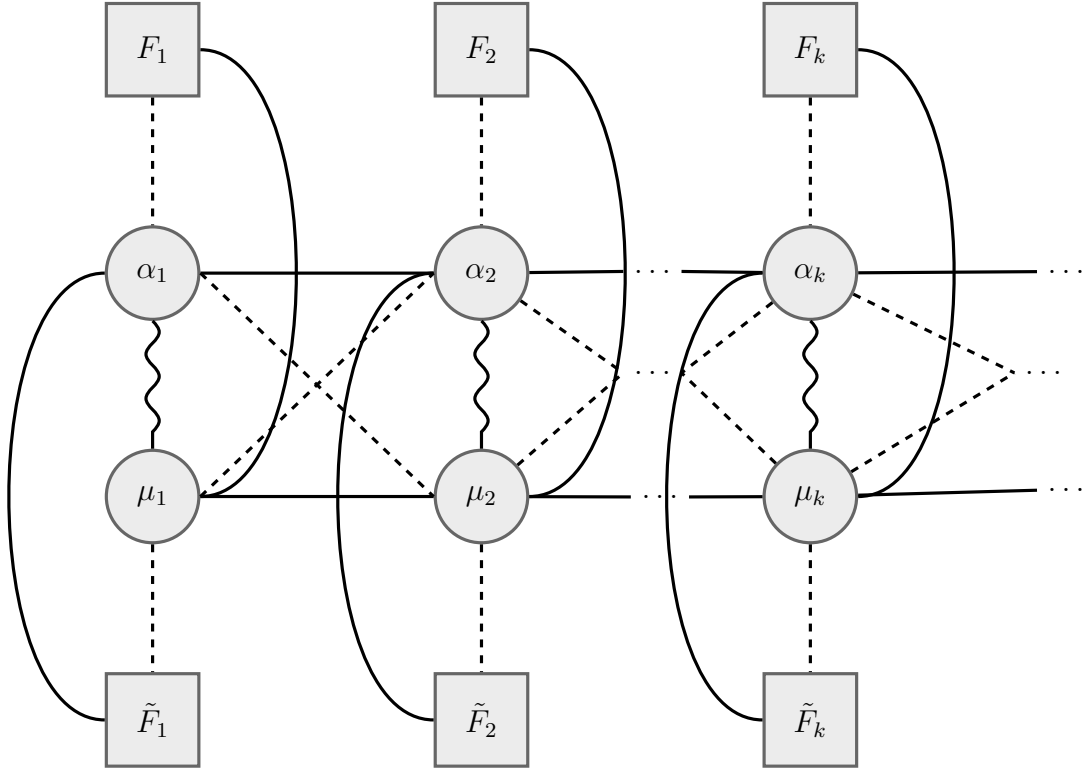
\begin{figure}[H]
\begin{center}
\begin{tikzpicture}
[
node distance = 17mm and 31mm,
b/.style={rectangle, draw=black!60, fill=gray!15, very thick, minimum size=35},
c/.style={circle, draw=black!60, fill=gray!15, very thick, minimum size=35}
]

\node[b]      (posFk)                            	{$F_k$};
\node[c]      (posak)      [below=of posFk]     	{$\alpha_k$};
\node[c]      (posmuk)     [below=of posak]      	{$\mu_k$};
\node[b]      (posFtk)     [below=of posmuk]      {$\tilde{F}_k$};

\node[b]      (posFF)      [left=of posFk]      	{$F_2$};
\node[c]      (posakk)     [below=of posFF]      	{$\alpha_2$};
\node[c]      (posmukk)    [below=of posakk]      {$\mu_2$};
\node[b]      (posFtkk)    [below=of posmukk]     {$\tilde{F}_2$};

\node[b]      (posFFF)     [left=of posFF]      	{$F_1$};
\node[c]      (posakkk)    [below=of posFFF]     	{$\alpha_1$};
\node[c]      (posmukkk)   [below=of posakkk]     {$\mu_1$};
\node[b]      (posFtkkk)   [below=of posmukkk]    {$\tilde{F}_1$};

\draw[-, very thick] (posmuk.east) 	.. controls  +(right:17mm) and +(right:17mm)   	.. (posFk.east);
\draw[-, very thick] (posak.west) 		.. controls  +(left:17mm) and +(left:17mm)   		.. (posFtk.west);
\draw[-, very thick] (posmukk.east) 	.. controls  +(right:17mm) and +(right:17mm)   	.. (posFF.east);
\draw[-, very thick] (posakk.west) 	.. controls  +(left:17mm) and +(left:17mm)   		.. (posFtkk.west);
\draw[-, very thick] (posmukkk.east) 	.. controls  +(right:17mm) and +(right:17mm)   	.. (posFFF.east);
\draw[-, very thick] (posakkk.west) 	.. controls  +(left:17mm) and +(left:17mm)   		.. (posFtkkk.west);

\draw[-, very thick, dashed] (posFk.south)  														to node[right] {} (posak.north);
\draw[-, very thick, decorate,decoration={coil,aspect=0,segment length=5.9mm}] (posak.south)  	to node[right] {} (posmuk.north);
\draw[-, very thick, dashed] (posmuk.south)  														to node[right] {} (posFtk.north);

\draw[-, very thick, dashed] (posFF.south)  														to node[right] {} (posakk.north);
\draw[-, very thick, decorate,decoration={coil,aspect=0,segment length=5.9mm}] (posakk.south)  	to node[right] {} (posmukk.north);
\draw[-, very thick, dashed] (posmukk.south)  													to node[right] {} (posFtkk.north);

\draw[-, very thick, dashed] (posFFF.south)  														to node[right] {} (posakkk.north);
\draw[-, very thick, decorate,decoration={coil,aspect=0,segment length=5.9mm}] (posakkk.south) 	to node[right] {} (posmukkk.north);
\draw[-, very thick, dashed] (posmukkk.south)  													to node[right] {} (posFtkkk.north);

\draw[-, very thick] (posakkk.east)  	to node[right] {} (posakk.west);
\draw[-, very thick] (posmukkk.east)  to node[right] {} (posmukk.west);

\draw[-, very thick, dashed] (posakkk.east)  	to node[right] {} (posmukk.west);
\draw[-, very thick, dashed] (posmukkk.east)  to node[right] {} (posakk.west);

\node at (31mm,-43mm) (aux0) {$\bf\textcolor{black}{\ldots}$};
\draw[-, very thick, dashed] (posak) to node[right] {} (aux0.west);
\draw[-, very thick, dashed] (posmuk) to node[right] {} (aux0.west);

\node at (34mm,-29.5mm) (aux1) {$\bf\textcolor{black}{\ldots}$};
\node at (34mm,-58.5mm) (aux2) {$\bf\textcolor{black}{\ldots}$};
\draw[-, very thick]  (posak) to node[right] {} (aux1);
\draw[-, very thick] (posmuk) to node[right] {} (aux2);

\node at (-21mm,-43mm) (aux3) {$\bf\textcolor{black}{\ldots}$};
\draw[-, very thick, dashed] (posakk) to node[right] {} (aux3.west);
\draw[-, very thick, dashed] (posmukk) to node[right] {} (aux3.west);
\draw[-, very thick, dashed] (posak) to node[left] {} (aux3.east);
\draw[-, very thick, dashed] (posmuk) to node[left] {} (aux3.east);
\node at (-21mm,-29.5mm) (aux4) {$\bf\textcolor{black}{\ldots}$};
\draw[-, very thick] (posakk) to node[right] {} (aux4.west);
\draw[-, very thick] (posak) to node[left] {} (aux4.east);
\node at (-20mm,-59.5mm) (aux5) {$\bf\textcolor{black}{\ldots}$};
\draw[-, very thick] (posmukk) to node[right] {} (aux5.west);
\draw[-, very thick] (posmuk) to node[left] {} (aux5.east);

\end{tikzpicture}
\end{center}
\caption{The general quiver with an IR limit that is captured by the $\text{AdS}_2$ background given by \cref{eq: ads2_geo,eq: ads2_nsns,eq: ads2_rr} and defined by \cref{eq: def_h4,eq: def_h8}. Each gauge node has vector multiplets as degrees of freedom. The straight solid and wavy lines represent the bifundamental hypermultiplets, while the dashed lines denote Fermi multiplets. In the seed 2d CFT the straight lines are (4,4) hypermultiplets, while the wavy lines (4,0).}
\label{fig: quiver}
\end{figure}

Here, we wish to discuss some basic aspects of the $\mathcal{N}=4$ superconformal quantum mechanical theories which were proposed to be the duals to the backgrounds that we consider in this work \cref{eq: ads2_geo,eq: ads2_nsns,eq: ads2_rr}. We have already mentioned that there deep relations between the $\text{AdS}_3$ and $\text{AdS}_2$ solutions, and as expected this is, also, reflected in the field theory side. The authors of \cite{Lozano:2020txg} argued that the type IIB solutions the obtained \cref{eq: ads2_geo,eq: ads2_nsns,eq: ads2_rr}, with the defining functions being given by \cref{eq: def_h4,eq: def_h8,eq: def_u} are dual to the IR limit of the quiver depicted in \cref{fig: quiver}. It is worthwhile stressing that the shape of this quiver diagram is directly inherited by the seed $\text{AdS}_3$ backgrounds \cite{Lozano:2020txg}. 

The proposal that was put forth in \cite{Lozano:2020txg} is that the way to think about the dynamics of the quantum mechanical quiver is to view it as the dimensional reduction along the spatial direction of the two-dimensional $\mathcal{N}=(4,0)$ quiver theories that were described in \cite{Lozano:2019zvg}. Since the shape of the quiver is inherited by the seed two-dimensional theory the same holds true for the matter content.

We can read off the ranks of colour and flavour groups from the Page charges. These were computed in \cite[equation 3.9]{Lozano:2020txg}. Let us discuss the $k^{th}$ entry of the quiver diagram that is in correspondence with the $[2\pi k, 2\pi (k+1)]$ interval of the geometry. We have $\text{U}(\alpha_k)$ and $\text{U}(\mu_k)$ for the colour groups and 
\begin{equation}
\alpha_k=\sum_{j=0}^{k-1}\beta_j		\,		,		\qquad	 \mu_k=\sum_{j=0}^{k-1}\nu_j		\,		. 
\end{equation}
These are coupled via bifundamental hypermultiplets and Fermi multiplets with the adjacent nodes. The connections with the $k^{th}$ flavour groups of ranks $\text{SU}(\nu_{k-1}-\nu_{k})$ and $\text{SU}(\beta_{k-1}-\beta_{k})$ is mediated by Fermi fields and by bifundamental hypermultiplets.

Finally, the various $F$ and $\tilde{F}$ are not independent of the other numbers of the quiver. It was noticed in \cite{Lozano:2019zvg} that the two-dimensional theory is chiral and might be afflicted by gauge anomalies. Hence, the $F$ and $\tilde{F}$ have to be chosen in such a way that gauge anomalies cancel out at each gauge node of the quiver. The exact relation was shown to be \cite{Lozano:2019zvg}
\begin{equation}
F_{k-1} = \nu_{k-1} - \nu_k \, , \quad \tilde{F}_{k-1} = \beta_{k-1} - \beta_k \, .
\end{equation}
	\subsubsection*{R-symmetry}
The R-symmetry group of supersymmetric $\mathcal{N} = 4$ quantum mechanics is $\text{SO}(4) = \text{SU}(2) \times \text{SU}(2)$. There are two possibilities for the superalgebra: the $\mathfrak{d}(2, 1; \alpha)$, with two $\mathfrak{su}(2)$ R-symmetries, and the $\mathfrak{su}(1, 1|2)$, with only one $\mathfrak{su}(2)$.

In the $\mathfrak{d}(2, 1; \alpha)$, which is usually named the large superconformal algebra, the parameter $\alpha$ is the one that parametrises the relative strength of the two Kac-Moody levels, $k_{-}$ and $k_{+}$ pertaining to the R-symmetries. Taking into consideration that we have only one $SU(2)$, the geometric realization being the one $\text{S}^2$ that appears in the backgrounds under consideration, we deduce that the superalgebra is the $\mathfrak{su}(1, 1|2)$. Another piece fo evidence comes from examining the seed $\text{AdS}_3$ backgrounds and the fact that the supercharges were in the $(\mathbf{1, 2;2})$ of $\text{SL}(2, \mathbb{R}) \times \text{SL}(2, \mathbb{R}) \times \text{SU}(2)_R$. 

We would, also, like to stress that the superalgebras in $d=1$ and $d=2$ dimensions are closely related to one another. More specifically, each chiral sector of a two-dimensional superconformal field theory provides a superalgebra and it is realisation for a $d=1$ superconformal quantum mechanics. This is what makes it possible to identify central charges in the cases of $d=1$ and $d=2$ \cite{Balasubramanian:2009bg} and it will important for the discussion of our results in \cref{sec: relation_ads3} where we compare the central charge that we compute in the type IIB $\text{AdS}_2$ solutions to the central charge of the seed $\text{AdS}_3$ theories. 
\section{Spin-2 fluctuations in the type IIB backgrounds}\label{sec: spin_2}
In order to begin our analysis we consider the background geometry given by \cref{eq: ads2_geo} in the Einstein frame. To do so, we multiply the string-frame metric, given by \cref{eq: ads2_geo}, by $e^{-\Phi/2}$, with $\Phi$ being the dilaton of our background. We can write it in the following manner
\begin{equation} \label{eq: geom_einstein_frame}
ds^2 = f_1 e^{- \Phi/2} ds^2_{\text{AdS}_2} + \hat{g}_{a b} dz^a dz^b \, ,
\end{equation}
with the warp factor and the internal part of the geometry being given by the following expressions:
\begin{equation}
	\begin{aligned}
f_1							&= \frac{1}{4}\frac{u}{\sqrt{\hat{h}_4 h_8}} \, , \\ 
\hat{g}_{a b} dz^a dz^b	&= e^{-\Phi/2} \left(  \frac{u \sqrt{\hat{h}_4 h_8}}{4 \hat{h}_4 h_8 + (u')^2} ds^2_{\text{S}^2} + \sqrt{\frac{\hat{h}_4}{h_8}} ds^2_{\text{CY}_2} + \frac{\sqrt{\hat{h}_4 h_8}}{u} (d \psi^2 + d \rho^2 ) \right) \, .
	\end{aligned}
\end{equation}
In what follows we will consider the Calabi-Yau to be a four-torus $\text{T}^4$ parametrised by four angles which we name $\theta_i, (i = 1, \dots , 4)$, and, of course, $\theta_i \cong \theta_i + 2 \pi$. We can consider a metric fluctuation, denoted by $h$, along the $\text{AdS}_2$ directions in the ten-dimensional background metric in the following manner:
\begin{equation}
ds^2 = f_1 e^{- \Phi/2} (ds^2_{\text{AdS}_2} + h_{\mu \nu} dx^{\mu} dx^{\nu}) + \hat{g}_{a b} dz^a dz^b \, .
\end{equation}
We can decompose $h$ into a transverse-traceless mode on $\text{AdS}_2$ and a scalar one along the internal manifold  like so:
\begin{equation}\label{eq: split_h}
h_{\mu \nu}(x,z) = h^{(tt)}_{\mu \nu}(x) \psi(z)		\,		. 
\end{equation}
Following the logic of \cite{Bachas:2011xa}, the transverse-traceless fluctuation $h^{(tt)}_{\mu \nu}(x)$ satisfies the following equation of motion on $AdS_2$
\begin{equation}\label{eq: ads_fluct}
\square_{\text{AdS}_2}^{(2)} h^{(tt)}_{\mu \nu}(x) = (M^2 -2) h^{(tt)}_{\mu \nu}(x) \, ,
\end{equation}
where $\square_{\text{AdS}_2}^{(2)}$ is the Laplacian acting on a massive rank-two tensor in $\text{AdS}_2$, see \cite{Polishchuk:1999nh}. In \cite{Bachas:2011xa}, the authors showed that the linearised Einstein equations can be reduced to the ten-dimensional Laplace equation given by:
\begin{equation}
\frac{1}{\sqrt{-g}} \partial_M \sqrt{-g} g^{MN} \partial_N h_{\mu \nu} = 0 \, .
\end{equation}
For the background metric in \cref{eq: geom_einstein_frame} and with the transverse-traceless $\text{AdS}_2$ mode, $h_{\mu \nu}^{(tt)}$, satisfying the equation \cref{eq: ads_fluct}, it is a matter of some straightforward algebra to obtain the following equation for the internal part of the fluctuation, $\psi(z)$,
\begin{equation}\label{eq: internal_eom}
\frac{1}{\sqrt{\hat{g}}} \partial_a \left[ f_1 e^{- \Phi/2} \sqrt{\hat{g}} \hat{g}^{ab} \partial_b \right] \psi = - M^2 \psi \, ,
\end{equation}
where we have suppressed the explicit dependence on $z$ for convenience.

We expand explicitly \cref{eq: internal_eom} to obtain
\begin{equation}\label{eq: internal_eom_expanded}
\left[ \frac{1}{\hat{h}_4 h_8} \frac{d}{d \rho} \left( u^2 \frac{d }{d \rho} \right) + \right( 4 + \frac{(u')^2}{\hat{h}_4 h_8} \left) \nabla^2_{S^2} + \frac{u}{\hat{h}_4} \sum^4_{i=1} \partial^2_{\theta_i} + \frac{u^2}{\hat{h}_4 h_8} \partial^2_{\psi} + M^2 \right] \psi = 0 \, .
\end{equation}

In order to proceed, it is convenient to decompose the mode $\psi$ into eigenstates along the different parts of the internal geometry. This can be done by considering the spherical harmonics on $\text{S}^2$, denoted by $Y_{\ell,m}$, and the plane-waves on the $\text{T}^4$ and $\mathcal{I}_{\psi}$ in the following manner:
\begin{equation}\label{eq: internal_decompose}
\psi = \sum \Psi_{\ell mnp} Y_{\ell,m} e^{i n \cdot \theta} e^{i p \cdot \psi} \, .
\end{equation}
In the above we have used $n$ as a short-hand for the four integers $(n_1, n_2, n_3, n_4)$ on the four-torus and $n \cdot \theta = n_1 \theta_1 + n_2 \theta_2 + n_3 \theta_3 + n_4 \theta_4$. Substituting \cref{eq: internal_decompose} into \cref{eq: internal_eom_expanded} we get the following equation for $\Psi$
\begin{equation}
	\begin{aligned}
\frac{d}{d\rho} \left(u^2 \frac{d \Psi_{\ell mnp}}{d \rho} \right) &- 4l(l+1) h_4 h_8 \Psi_{\ell mnp} - (u^{\prime})^2 l (l+1) \Psi_{\ell mnp} \\
&- \frac{u}{h_4}h_4 h_8 n^2 \Psi_{\ell mnp} - u^2 p^2 \Psi_{\ell mnp} + 4 h_4 h_8 M^2 \Psi_{\ell mnp} = 0		\,			.
	\end{aligned}
\end{equation}
It is convenient to consider $\Psi_{\ell mnp} = u^{\ell} \varphi$ in the above equation. Of course, $\varphi$ also depends on all the quantum numbers and we should formally write $\varphi = \varphi_{\ell mnp}$, however, we suppress the subscripts in what follows. In terms of $\varphi$ we obtain:
\begin{equation}\label{eq: help_de}
	\begin{aligned}
u^{\ell+2} \frac{d \varphi}{d\rho} &+ 2\ell(\ell+1) \frac{d\varphi}{d\rho}\frac{d u}{d\rho} \\
&- 4\ell(\ell+1)h_4 h_8 u^{\ell} \varphi - \frac{u}{h_4} h_4 h_8 n^2 u^{\ell} \varphi - u^2 p^2 u^{\ell} \varphi + 4 h_4 h_8 M^2 u^{\ell} \varphi = 0			\,			. 
	\end{aligned}
\end{equation}
In deriving the above, we used $u^{\prime \prime}=0$ which holds globally true.  

We can re-express \cref{eq: help_de} in a more suggestive and intuitive form, namely we can write it as a Sturm-Liouville problem\footnote{the integrating factor in this case turns out to be given by $\mu(\rho)=u^{\ell}$}
\begin{equation}\label{eq: sturm_liouville_1}
S(\rho) \varphi + Q(\rho) \varphi = - \lambda W(\rho) \varphi			\,			,
\end{equation}
with 
\begin{gather}\label{eq: sturm_liouville_2}
    \begin{aligned}
    S(\rho) &= \frac{d}{d\rho}\left(P(\rho) \frac{d}{d\rho} \right)	\,	,
    & Q(\rho) &= - (n^2 h_8 u + p^2 u^2)u^{2 \ell} \varphi				\,	,
    \\
    P(\rho) &= u^{2(\ell+1)}												\,	,
    & W(\rho) &= h_4 h_8 u^{2\ell}										\,	,
    \end{aligned}
\end{gather}
and the eigenvalue $\lambda$ being given by 
\begin{equation}\label{eq: sturm_liouville_3}
\lambda = 4 (M^2 - \ell(\ell+1))		\,		.
\end{equation}

The variable $\rho$ takes values in the $\mathcal{I}_{\rho}$ interval; $[0,2\pi(P+1)]$. The function $u$ vanishes at the $\rho=0$ endpoint, and hence we have a singular Sturm-Liouville problem. 
\section{Unitarity bound and universal class of solutions}\label{sec: universal}
We would like to proceed by obtaining a bound for $M^2$ that emerges from the Sturm-Liouville problem derived in the previous section, \cref{eq: sturm_liouville_1,eq: sturm_liouville_2,eq: sturm_liouville_3}. When this bound is saturated we are able to find a specific class of solutions which we will call minimal universal class. This class of solutions is independent of the defining functions $\hat{h}_4$ and $h_8$. We discuss the regularity conditions for the scalar mode $\Psi$ as well.
	\subsection*{The unitarity bound}
We begin by multiplying the Sturm-Liouville problem defined by \cref{eq: sturm_liouville_1,eq: sturm_liouville_2,eq: sturm_liouville_3} through by $\varphi$ and then we carry out the integration over $\mathcal{I}_{\rho}$ interval. That is, we obtain 
\begin{equation}
\int_{\mathcal{I}_{\rho}} d \rho \, \varphi \frac{d}{d \rho} \left( u^{2(\ell+1)} \frac{d \varphi}{d \rho} \right) - \left( n^2 h_8 u + p^2 u^2 \right) u^{2 \ell} \varphi^2 + 4 \left( M^2 - \ell (\ell+1) \right) \hat{h}_4 h_8 u^{2 \ell} \varphi^2 = 0 \, ,
\end{equation}
We focus on the first term of the above and we perform an integration by parts. This gives
\begin{equation}\label{eq: def_bound}
\begin{split}
\int_{\mathcal{I}_{\rho}} d \rho \left( - \varphi'^2  u^{2(\ell+1)} - \left( n^2 h_8 u + p^2 u^2 \right) u^{2 \ell} \varphi^2 + 4 \left( M^2 - \ell (\ell+1) \right) \hat{h}_4 h_8 u^{2 l} \varphi^2 \right) = \\
- \varphi \varphi' u^{2(\ell+1)}\Big\rvert_{0}^{\rho^{*}} \, .
\end{split}
\end{equation}
We note that the right-hand side of \cref{eq: def_bound}, $\varphi \varphi' u^{2(\ell+1)}$, vanishes when we evaluate it at the endpoint $\rho = 0$, since $u$ vanishes at $\rho = 0$ - recall that $u$ is linear in $\rho$ and its exact definition is given by \cref{eq: def_u} - as long as $\varphi$ and $\varphi'$ are regular there, while it does not vanish when evaluated at the other endpoint; $\rho = \rho^*$. In what follows, we will focus on functions $\varphi$ such that which $\varphi \varphi' u^{2(\ell+1)}$ vanishes at $\rho = \rho^*$ as well. Therefore, \cref{eq: def_bound} reduces to the following:
\begin{equation} \label{eq: for_bound}
\int_{\mathcal{I}_{\rho}} d \rho \left(  \varphi'^2  u^{2(\ell+1)} + \left( n^2 h_8 u + p^2 u^2 \right) u^{2\ell} \varphi^2 \right) = 4 (M^2 - \ell (\ell+1)) \int_{\mathcal{I}_{\rho}} d \rho \, \hat{h}_4 h_8 u^{2 \ell} \varphi^2 \, ,
\end{equation}
Taking into consideration that the defining functions of our backgrounds $u$, $\hat{h}_4$ and $h_8$ are non-negative, and that the integrals are finite, we obtain from \cref{eq: for_bound} the following lower bound for $M^2$
\begin{equation}\label{eq: bound}
M^2 \geq \ell(\ell+1) \, .
\end{equation}
	\subsection*{Minimal universal class of solutions}
Let us consider here the case where $M$ saturates its bound; $M^2 = \ell(\ell+1)$. We, also, consider the case $n = p = 0$, or in other words the modes are not excited along the directions of the $\text{T}^4$ and the $\mathcal{I}_{\psi}$. After those considerations, the Sturm-Liouville problem defined by \cref{eq: sturm_liouville_1,eq: sturm_liouville_2,eq: sturm_liouville_3} reduces to
\begin{equation}\label{eq: universal_minimal}
\frac{d}{d \rho} \left( u^{2(\ell+1)} \frac{d \varphi}{d \rho} \right) = 0 \, .
\end{equation}
We can integrate \cref{eq: universal_minimal} to obtain 
\begin{equation}
\varphi^{\prime} = \frac{\texttt{constant}}{u^{2(\ell+1)}}				\,		.
\end{equation}

However, for the class of geometries considered in this work, $u$ vanishes at $\rho = 0$, and therefore $\varphi^{\prime}$ is not finite at $\rho = 0$. This suggests that both $\varphi$ and $\Psi_{\ell m} = u^{\ell} \varphi$ are not finite at $\rho = 0$. However, we are interested in finding fluctuations for the modes that remain finite anywhere in the allowed region of the geometries that we consider, and hence the only solution to \cref{eq: universal_minimal} that satisfies this condition is the constant solution. This, in turn, implies that
\begin{equation}\label{eq: universal_sltn}
\varphi = \texttt{constant} 	\, , \qquad \Psi = \texttt{constant} \cdot u^{\ell} 	\, , \qquad M^2 = \ell(\ell+1) \, .
\end{equation}
This class of solutions does not depend on the form that the functions $u$, $\hat{h}_4$ and $h_8$ have and in this sense it is a universal class of solutions. Furthermore, they are the solutions obtained for the minimal value of $M$ for a given $\ell$.

More specifically, the spin-2 fluctuations that we considered above are dual to field theory operators that carry dimension $\Delta$. The conformal dimensions of those operators is related to the mass via $M^2 = \Delta(\Delta - 1)$ using the standard AdS/CFT formula. The inequality derived earlier, \cref{eq: bound}, implies for the conformal dimension of the field theory operators is bounded from below
\begin{equation}\label{eq:bound on the dimension}
\Delta \geq \Delta_{\texttt{min}} \, ,  \qquad \Delta_{\texttt{min}} = \ell + 1 \, .
\end{equation}

For the case of non-universal solutions, i.e. solutions for which $M^2 > \ell(\ell+1)$ holds, or non-minimal solutions, such that the modes are excited on $\text{T}^4$ or along the $\mathcal{I}_{\psi}$, it is necessary to specify what the three functions $u$, $\hat{h}_4$ and $h_8$ are.
\section{The holographic central charge}\label{sec: central_charge}
In this section we will compute the central charge for the theories described by \cref{eq: ads2_geo,eq: ads2_rr,eq: ads2_nsns}. Before we proceed with our computation, some commentary is in order. Trying to define the central charge in conformal quantum mechanics is a very subtle issue. The reasoning is as follows: let us consider a one-dimensional conformal theory. Then, we have only one component of the stress-energy tensor, $T_{\mu\nu}$, the trace of which must vanish, and hence this implies that $T_{tt}=0$. The quantity that we name as holographic central charge should be interpreted as the number of vacuum states in the associated superconformal quantum mechanics picture \cite{Lozano:2020txg}. Our reasoning behind choosing that particular name is to stay consistent with the existing literature.

We will be following the approach described \cite{Gutperle:2018wuk}. The central charge has been, also, computed in \cite{Lozano:2020txg} using the algorithmic approach developed first in \cite{Klebanov:2007ws} and revisited in \cite{Macpherson:2014eza} to account for more general warp factors.. 

Our starting point is the action for the type IIB theory, which we only need schematically. In the Einstein frame it reads 
\begin{equation}
S_{\text{IIB}} = \frac{1}{2 \kappa_{10}^2} \int d^{10} x \sqrt{-g} \left( R + \cdots \right)		\, .
\end{equation}
Following the logic outlined in \cite{Gutperle:2018wuk}, we expand the supergravity action to second order and we are led to an action for the $h_{\mu \nu}$
\begin{equation}\label{eq: action_for_h}
\delta^2 S = \frac{1}{\kappa_{10}^2} \int d^{10}x \, {h}^{\mu \nu} \partial_M \sqrt{-g} g^{MN} \partial_{N} h_{\mu \nu} + \texttt{boundary terms} \, .
\end{equation}
We can evaluate \cref{eq: action_for_h} for the bulk and internal pieces and drop the boundary terms to obtain
\begin{equation}\label{eq: action_h_expanded}
S[h] = \frac{1}{\kappa_{10}^2} \int d^{10}x \sqrt{-g_{AdS_2}} \sqrt{\hat{g}} \, {h}^{\mu \nu} \left\{ \square_{AdS_2}^{(2)} + 2 + \hat{\square} \right\} h_{\mu \nu} \, ,
\end{equation}
with $\hat{\square}$ in the above being the operator on the left-hand side of \cref{eq: internal_eom}. We decompose the metric perturbations by using the ansatz
\begin{equation}
h_{\mu \nu} = (h_{lmnp}^{(tt)})_{\mu \nu} Y_{lm} \psi_{lmnp} e^{i n \cdot \theta} e^{i n \cdot \theta} e^{i p \cdot \psi}  
\end{equation}
and we find
\begin{equation}
\delta^2 S = \sum_{lmnp} C_{lmnp} \int d^2 x  \sqrt{-g_{AdS_2}} \, {(h^{(tt)}_{lmnp})}^{\mu \nu} \left\{ \square_{AdS_2}^{(2)} + 2 - M^2 \right\} {(h^{(tt)}_{lmnp})}_{\mu \nu}
\end{equation}
where the coefficients $C_{lmnp}$ are given by
\begin{equation}\label{eq: normalisation_spin2}
C_{lmnp} = \frac{1}{\kappa_{10}} \vol_{T^4} \vol_{S^2} \vol_{\psi} \int_{\mathcal{I}_{\rho}} d \rho \sqrt{\hat{g}} \,  |\psi_{lmnp}|^2 \, ,
\end{equation}
and we have used the standard normalisation $\int Y_{lm} Y_{l' m'} = \delta_{l l'} \delta_{m m'}$ for the spherical harmonics.

The integral in \cref{eq: normalisation_spin2} is finite for the all those solutions discussed described by fluctuations that are finite everywhere. In particular, by specialising to the minimal universal solution class of solution, \cref{eq: universal_sltn}, and by setting all the quantum numbers to zero, $l=m=n=p=0$ such that $\psi_{lmnp} = 1$, \cref{eq: universal_sltn} evaluated on \cref{eq: ads2_geo} defined in the Einstein frame yields 
\begin{equation}\label{eq: ct}
C_{\{0\}} = \frac{1}{4 \kappa_{10}^2} \vol_{T^4} \vol_{S^2} \vol_{\psi} \int_{\mathcal{I}_{\rho}} d \rho \, \hat{h}_4 h_8 \, .
\end{equation}
For the volume-forms in the above we use 
\begin{equation}
\vol_{T^4} = 16 \pi^4	\,	,	\qquad 	\vol_{S^2} = 4 \pi	\,	,	\qquad	 \vol_{\psi} = 2 \pi	\,	,
\end{equation}
and also 
\begin{equation}
2 \kappa^2 = (2\pi)^7	\,	,	
\end{equation}
in units such that $\alpha^{\prime} = g_s = 1$ in order to follow closely \cite{Lozano:2020txg}. Then, the final result is 
\begin{equation}\label{eq: ct_final}
C_{\{0\}} = \frac{1}{2 \pi} \int_{\mathcal{I}_{\rho}} d \rho \, \hat{h}_4 h_8 \, ,
\end{equation}
which agrees with \cite[equation 4.4]{Lozano:2020txg} modulo an irrelevant numerical factor. 

In fact, if we are persistent on fixing the final numerical factor we can consider the integral along the AdS directions. The integrand is zero, which in turn implies that we have the multiplication of \cref{eq: ct_final} by a constant. That constant can be fixed by matching directly to the holographic central charge computed in \cite[equation 4.4]{Lozano:2020txg}. 
\section{On the SU(1,1\texorpdfstring{$|$}{|}2) multiplets}\label{sec: multiplets}
For each one of the mode solutions that are derived from the fluctuation of the metric that we have considered in this work, there exists an operator in the dual field theory picture. We should attempt to understand to what kind of operators these metric perturbations correspond. With that in mind, it will be useful to try and identify the associated field theory multiplets. We will denote the states that correspond to the metric fluctuations using the relevant conformal dimension and the angular quantum number due to the $\text{S}^2$ isometries. Hence we have 
\begin{equation}
[\Delta,\ell]		\,		.
\end{equation}

The unitary, irreducible representations of the $\text{SU}(1,1|2)$ have been constructed long ago in \cite{Gunaydin:1986fe}, see also \cite{Aharony:1999ti}. The short multiplets are labelled by a half-integer
\begin{equation}
(k,k) \oplus 2(k+\tfrac{1}{2},k-\tfrac{1}{2}) \oplus (k+1,k-1)	\,	,
\end{equation}
and we have in total four states, the $\text{SL}(2,\mathbb{R})$ primaries, except for the lowest case, $k = \tfrac{1}{2}$, where the last one is absent.

As we pointed out earlier, the mass of a spin-2 field in the bulk and the conformal dimension of the dual operator are related by means of $M^2=\Delta(\Delta-1)$. Therefore, the minimal universal solution derived earlier, \cref{eq: universal_sltn}, for which we have $M^2=\ell(\ell + 1)$, corresponds to operators with scaling dimension $\Delta = \ell + 1$. Therefore we can re-write the state as 
\begin{equation}
[\ell+1,\ell]		\,		.
\end{equation}
Starting from the lowest value of the $\text{S}^2$ harmonics $\ell=0$, we that the state is $[1,0]$. Taking into consideration that this should correspond to the $k=\tfrac{1}{2}$ value we can relate the values as $k=\tfrac{1}{2}+\ell$. 
\section{Non-universal classes of solutions}\label{sec: non_uni}
Here, we wish to consider two particular classes of solutions to the Sturm-Liouville problem, which are distinct from the universal ones. In other words, the solutions we discuss here do not saturate the bound $M^2 = \ell(\ell+1)$.
	\subsection*{Case I: Solutions with \texorpdfstring{$n=p=0$}{np0}}
We start by considering $n=p=0$ still. This means that the bulk modes are not excited along the $\text{T}^4$ directions and on the $S^1$. The difference with the minimal universal class is only that the mass does not saturate the unitarity bound. In order to solve the Sturm-Liouville problem, we will have to choose some particular $u$, $\hat{h}_4$ and $h_8$. This choice, in turn, corresponds to a particular background. 

Let us start off by considering the case of
\begin{equation} \label{eq: special_sugra_def}
\hat{h}_4 = h_8 = \beta_0 \begin{cases} 
      \rho/2 \pi & 0 \leq \rho \leq  \pi P\\
      P - \rho/2 \pi &   \pi P < \rho \leq 2 \pi P
   \end{cases} \, , 
    \quad \quad u = \frac{\beta_0}{2 \pi} \rho \, .
\end{equation}
A solution to \cref{eq: sturm_liouville_1,eq: sturm_liouville_2,eq: sturm_liouville_3} splits into two solutions along the two intervals $ \mathcal{I}_1 = \left[0,  \pi P \right]$ and $\mathcal{I}_2 = \left(\pi P, 2 \pi P \right]$, since both $\hat{h}_4$ and $h_8$ are only piecewise continuous. Moreover, since we are interested in a smooth solution for the fluctuations, we will impose that the solution and its derivative are continuous at $\rho = \pi P$.

For the particular choice of $u$, $\hat{h}_4$, $h_8$ we made here and for $n=p=0$ the Sturm-Liouville problem \cref{eq: sturm_liouville_1,eq: sturm_liouville_2,eq: sturm_liouville_3} becomes in $\mathcal{I}_1$
\begin{equation}\label{eq: eom_non_uni_1}
\varphi''(\rho) + \frac{2 \ell + 2}{\rho} \varphi'(\rho) + \lambda ~ \varphi(\rho) = 0  \,	,	
\end{equation}
while in $\mathcal{I}_2$ we have 
\begin{equation}\label{eq: eom_non_uni_2}
\varphi''(\rho) + \frac{2 \ell + 2}{\rho} \varphi'(\rho) + \lambda ~ \frac{ (P-\rho/2 \pi)^2}{\rho^2} \varphi(\rho) = 0 \,		,
\end{equation}
with $\lambda$ being given by \cref{eq: sturm_liouville_3}. The general solution of \cref{eq: eom_non_uni_1} reads
\begin{equation}\label{eq: sltn_i1_1}
\varphi = \rho^{-\ell-\frac{1}{2}} \left( c_1~J_{\ell + \frac{1}{2}} \big( \sqrt{\lambda} \rho \big) + c_2~Y_{\ell + \frac{1}{2}} \big( \sqrt{\lambda} \rho  \big)  \right) \, ,
\end{equation}
with $J_n(z)$ and $Y_n(z)$ being the Bessel functions of the first and second kind, respectively, and $c_{1,2}$ are just the integration constants. However, since we are interested in the solution and its derivative anywhere within $\mathcal{I}_1$ we must set $c_2 = 0$ and hence the final solution reads: 
\begin{equation}\label{eq: sltn_non_uni_phi_1}
\varphi = c_1~\rho^{-\ell-\frac{1}{2}}~J_{\ell + \frac{1}{2}} \big( \sqrt{\lambda} \rho \big)  \, .
\end{equation}
We can, also, solve \cref{eq: eom_non_uni_2} whose general solution is given by:
\begin{equation}\label{eq: sltn_i1_2}
\varphi = e^{- \sqrt{\lambda} \rho} \rho^{-(\ell+\frac{1}{2}-\frac{\gamma}{2})} \left(c_3 ~ U(\alpha,1+\gamma,2i\sqrt{\lambda}\rho) + c_4 ~ L^{\gamma}_{-\alpha}(2i\sqrt{\lambda}\rho) \right)	\,	,
\end{equation}
where in the above $U(a,b,z)$ is the confluent hypergeometric function, $L^{a}_b(z)$ is the generalized Laguerre polynomial, $c_{3,4}$ are integration constants and we have defined the auxiliary parameters $\alpha, \gamma$ as follows: 
\begin{equation}
	\begin{aligned}
\alpha &= \frac{1}{2} - 2i\pi P \sqrt{\lambda} + \frac{1}{2}\gamma	\,	,	\\
\gamma &= \sqrt{(1+2\ell)^2-16P^2\pi^2 \lambda}	\,	.
	\end{aligned}
\end{equation}
The above solution is well-defined everywhere along the $\mathcal{I}_2$ interval and hence there is no need to tune any of the constants to zero. 

We can now proceed to match the solutions derived above and their derivatives at $\rho = \pi P$. This leads to two conditions. A third condition is coming from imposing that the solution we derived in the $\mathcal{I}_2$ interval or its derivative should vanish at $\rho = 2 \pi P$; see the discussion below \cref{eq: def_bound}. We choose to make the function vanish. After we take these into consideration, we have three equations and we need to determine three unknown parameters, namely $c_{1,3,4}$. This system admits a non-trivial solution if and only if the determinant of the following matrix
\begin{equation}
\begin{pmatrix} 
\varphi_1(\rho = \pi P) 				& - \varphi_3(\rho = \pi P)					& -  \varphi_4(\rho = \pi P) \\
\varphi^{\prime}_1(\rho = \pi P) 		& - \varphi^{\prime}_3(\rho = \pi P) 			& - \varphi^{\prime}_4(\rho = \pi P) \\
0 										& \varphi_3(\rho = 2 \pi P) 					&  \varphi_4(\rho = 2 \pi P) 
\end{pmatrix} 		\, 	,
\end{equation}
is equal to zero. 

It is at this point that we would have to resort to a careful numerical analysis for different $M^2$ such that the above holds true. The expectation is that the consistent solution for $M^2$ is of the schematic form
\begin{equation}\label{eq: non_universal_mass_relation}
M^2 = \ell (\ell+1) + p_1 f(p_1, p_2) \, , 
\end{equation}
with $p_1$ would be a positive integer number and $f$ a positive function such that $f(0, p_2)$ is a regular solution. 

It is possible realize and explain the form of \cref{eq: non_universal_mass_relation} analytically by examining carefully the results and conclusions of \cref{sec: universal}. We know from the holographic dictionary that $M^2 = \Delta(\Delta-1)$ in the case of AdS$_2$/CFT$_1$. We, also, know that the conformal dimension is bounded from below owing to the unitarity of the CFT, in other words $\Delta \geq \Delta_{\texttt{min}}$ with $\Delta_{\texttt{min}} = \ell + 1$. The value of $\Delta_{\texttt{min}}$ is carried by the universal spin-2 fluctuations that were studied in \cref{sec: universal} and are present in all the CFT duals of these AdS$_2$ backgrounds, regardless of the form of the defining functions $u, \hat{h}_4$ and $h_8$. These universal fluctuations are, of course, present in the specific examples that we study in this section. The field theory operators dual to the modes described by \cref{eq: sltn_non_uni_phi_1,eq: sltn_i1_2} necessarily carry a conformal dimension bigger than that of $\ell + 1$. This is reflected in their mass spectrum by the addition of $p_1 f(p_1, p_2)$\footnote{in simpler terms this quantity denotes a numerical interpolating function.} to the minimum $M^2$ that was derived in \cref{eq: universal_sltn}. 
	\subsection*{Case II: Solutions with \texorpdfstring{$n=0$}{n0}}
Here, we allow the modes to have some excitation on the $\text{S}^1$ while keeping them at $n=0$, i.e no excitation along the $\text{T}^4$. We choose to examine, as before, the backgrounds described by defining functions given by \cref{eq: special_sugra_def}. Similarly to our previous consideration, the Sturm-Liouville problem \cref{eq: sturm_liouville_1,eq: sturm_liouville_2,eq: sturm_liouville_3}  is split into two part. In $\mathcal{I}_1$ interval the equation that governs the spin-2 dynamics is given by:
\begin{equation}\label{eq: eom_non_uni_1_b}
\varphi''(\rho) + \frac{2 \ell + 2}{\rho} \varphi'(\rho) + (\lambda - p^2) ~ \varphi(\rho) = 0  \,	,	
\end{equation}
while in $\mathcal{I}_2$ we have 
\begin{equation}\label{eq: eom_non_uni_2_b}
\varphi''(\rho) + \frac{2 \ell + 2}{\rho} \varphi'(\rho) + \left(-p^2 + \lambda ~ \frac{ (P-\rho/2 \pi)^2}{\rho^2} \right) \varphi(\rho) = 0 \,		,
\end{equation}
with $\lambda$ being given by \cref{eq: sturm_liouville_3}. 

It is obvious that the resulting equations in this case, \cref{eq: eom_non_uni_1_b,eq: eom_non_uni_2_b}, are very similar to the ones obtained when we did not consider any excitation on the $\text{S}^1$, see \cref{eq: eom_non_uni_1,eq: eom_non_uni_2}. The solutions to the above equations are the ones we expect from the previous case. Namely, the general solution of \cref{eq: eom_non_uni_1_b} reads
\begin{equation}
\varphi = \rho^{-\ell-\frac{1}{2}} \left( c_1~J_{\ell + \frac{1}{2}} \big(-i\sqrt{p^2-\lambda} \rho \big) + c_2~Y_{\ell + \frac{1}{2}} \big(-i\sqrt{p^2-\lambda} \rho  \big)  \right) \, ,
\end{equation}
with $J_n(z)$ and $Y_n(z)$ being the Bessel functions of the first and second kind, respectively, and $c_{1,2}$ are just the integration constants. However, since we are interested in the solution and its derivative anywhere within $\mathcal{I}_1$ we must set $c_2 = 0$ and hence the final solution reads: 
\begin{equation}
\varphi = c_1~\rho^{-\ell-\frac{1}{2}}~J_{\ell + \frac{1}{2}} \big( -i\sqrt{p^2-\lambda} \rho \big)  \, .
\end{equation} 
We can, also, solve \cref{eq: eom_non_uni_2} whose general solution is given by:
\begin{equation}
\varphi = e^{- \sqrt{p^2-\lambda} \rho} \rho^{-(\ell+\frac{1}{2}-\frac{\gamma}{2})} \left(c_3 ~ U(\alpha,1+\gamma,2\sqrt{p^2-\lambda}\rho) + c_4 ~ L^{\gamma}_{-\alpha}(2\sqrt{p^2-\lambda}\rho) \right)	\,	,
\end{equation}
where in the above $U(a,b,z)$ is the confluent hypergeometric function, $L^{a}_b(z)$ is the generalized Laguerre polynomial, $c_{3,4}$ are integration constants and we have defined the auxiliary parameters $\alpha, \gamma$ as follows: 
\begin{equation}
	\begin{aligned}
\alpha &= \frac{1}{2} + \frac{2\pi P \lambda}{\sqrt{p^2 - \lambda}} + \frac{1}{2}\gamma	\,	,	\\
\gamma &= \sqrt{(1+2\ell)^2-16P^2\pi^2 \lambda}	\,	.
	\end{aligned}
\end{equation}
The above solution is well-defined everywhere along the $\mathcal{I}_2$ interval and hence there is no need to tune any of the constants to zero. 

Of course, one would have to proceed as before in order to set up a system of equations that would need to analyse numerically, with very minor changes from our previous treatment. More specifically, we again proceed to match the solutions derived above and their derivatives at the point $\rho = \pi P$. This leads to two conditions. A third condition is coming from imposing that the solution we derived in the $\mathcal{I}_2$ interval or its derivative should vanish at $\rho = 2 \pi P$; see the discussion below \cref{eq: def_bound}. We choose to make the function vanish. After we take these into consideration, we have three equations and we need to determine three unknown parameters, namely $c_{1,3,4}$. This system admits a non-trivial solution if and only if the determinant of the following matrix
\begin{equation}
\begin{pmatrix} 
\varphi_1(\rho = \pi P) 				& - \varphi_3(\rho = \pi P)					& -  \varphi_4(\rho = \pi P) \\
\varphi^{\prime}_1(\rho = \pi P) 		& - \varphi^{\prime}_3(\rho = \pi P) 			& - \varphi^{\prime}_4(\rho = \pi P) \\
0 										& \varphi_3(\rho = 2 \pi P) 					&  \varphi_4(\rho = 2 \pi P) 
\end{pmatrix} 		\, 	,
\end{equation}
is equal to zero. 

The mass spectrum in this case, also, assumes the generic form
\begin{equation}\label{eq: non_universal_mass_relation_ii}
M^2 = \ell (\ell+1) + r_1 g(r_1, r_2) \, , 
\end{equation}
with $r_1$ would be a positive integer number and $g$ a positive function such that $g(0, r_2)$ is a regular solution. 

It is possible realize and explain the form of \cref{eq: non_universal_mass_relation_ii} analytically by examining carefully the results and conclusions of \cref{sec: universal}. We know from the holographic dictionary that $M^2 = \Delta(\Delta-1)$ in the case of AdS$_2$/CFT$_1$. We, also, know that the conformal dimension is bounded from below owing to the unitarity of the CFT, in other words $\Delta \geq \Delta_{\texttt{min}}$ with $\Delta_{\texttt{min}} = \ell + 1$. The value of $\Delta_{\texttt{min}}$ is carried by the universal spin-2 fluctuations that were studied in \cref{sec: universal} and are present in all the CFT duals of these AdS$_2$ backgrounds, regardless of the form of the defining functions $u, \hat{h}_4$ and $h_8$. These universal fluctuations are, of course, present in the specific examples that we study in this section. The field theory operators dual to the modes described by \cref{eq: sltn_non_uni_phi_1,eq: sltn_i1_2} necessarily carry a conformal dimension bigger than that of $\ell + 1$. This is reflected in their mass spectrum by the addition of $r_1 g(r_1, r_2)$\footnote{in this case, as in the previous one, it is just a numerical interpolating function.} to the minimum $M^2$ that was derived in \cref{eq: universal_sltn}.
\section{From \texorpdfstring{$\text{AdS}_2/\text{CFT}_1$}{ads2cft1} to \texorpdfstring{$\text{AdS}_3/\text{CFT}_2$}{ads3cft2}}\label{sec: relation_ads3}
In this section we discuss how we can use the $\text{AdS}_2/\text{CFT}_1$ picture developed in \cite{Lozano:2020txg} to extract information about the seed $\text{AdS}_3/\text{CFT}_2$ theories \cite{Lozano:2019emq}. 
	\subsection*{The spin-2 dynamics}
Let us recall the basic equation that governs the dynamics of the spin-2 modes before decomposing along the various submanifolds
\begin{equation}\label{eq: internal_eom_expanded_2}
\left[ \frac{1}{\hat{h}_4 h_8} \frac{d}{d \rho} \left( u^2 \frac{d }{d \rho} \right) + \right( 4 + \frac{(u')^2}{\hat{h}_4 h_8} \left) \nabla^2_{S^2} + \frac{u}{\hat{h}_4} \sum^4_{i=1} \partial^2_{\theta_i} + \frac{u^2}{\hat{h}_4 h_8} \partial^2_{\psi} + M^2 \right] \psi = 0 \, .
\end{equation}
The difference between the above and \cite[equation (16)]{Speziali:2019uzn} is merely the $\partial^2_{\psi}$-term which is associated with the $\text{S}^1$. However, keeping in mind that the $\text{S}^1$ was derived by a T-duality transformation along the spatial direction of the seed $\text{AdS}_3$, one can just disregard it if the interest is to extract information about the seed backgrounds from the lower-dimensional ones. 
	\subsection*{The minimal universal solutions}
The minimal universal solutions described by \cref{eq: universal_sltn} are exactly the same as those obtained in the seed $\text{AdS}_3$ backgrounds, see \cite[equation (28)]{Speziali:2019uzn}. This is perhaps not too surprising, since there are many structural similarities between the two different backgrounds as they are T-duals. Of course, the unitarity bounds are different in the two cases, however, by adopting the line of reasoning described above and solving the modified $\text{AdS}_2$ Sturm-Liouville problem one would be able to start from an $\text{AdS}_2$ study and derive a result pertaining to the two-dimensional superconformal field theory dual to the $\text{AdS}_3$ background solutions.
	\subsection*{The non-universal solutions}
The reason that we chose the defining functions to have the particular form given by \cref{eq: special_sugra_def} was to make a direct comparison \cite[appendix A]{Speziali:2019uzn}. Specifically, the non-universal Class I mode solutions are the same both in the $\text{AdS}_2$ backgrounds and the $\text{AdS}_3$ solutions. We are referring to our solutions given by \cref{eq: sltn_i1_1,eq: sltn_i1_2} that can be compared directly to \cite[equations (39) and (40)]{Speziali:2019uzn}. A more interesting question would be, perhaps, to try and understand how the Class II non-universal solutions are related to the $\text{AdS}_3$ backgrounds. Recall, that these are the solutions obtained by considering some non-trivial excitation on the $\text{S}^1$ which is obtained by a T-duality transformation along an AdS isometry in the seed massive type IIA theories.  
	\subsection*{The holographic central charge}
A final comment needs to be made before we close this section. From the R-symmetry discussion in \cref{sec: sugra} we expect that there is a match between the central charge we computed here, and the central charge of the $\mathcal{N}=(4,0)$ $\text{AdS}_3$ seed solutions. In the latter context, the effective three-dimensional gravitational coupling $\kappa_3$ is related to $C_{\{0\}}$ by  $C_{\{0\}} = 1/\kappa_3^2$. Also, the expanded quadratic action for the $h_{\mu \nu}$ fluctuations computes the two-point function of the dual stress-energy tensor. The coefficient of the latter is proportional to the central charge of the dual two-dimensional superconformal field theory. \Cref{eq: ct_final} is, as expected, equal to the central charge derived in \cite[equation 2.27]{Lozano:2019zvg} up to an unimportant numerical factor, which can be fixed as we described earlier in \cref{sec: central_charge}.

A, perhaps, more natural and straightforward comparison for the expression of the central charge can be made when considering the result we derived earlier in \cref{eq: ct_final} and the result for the central charge computed in \cite{Speziali:2019uzn}. In the latter, the same method that was adopted here was used to compute the central charge of the seed field theory solutions, namely the spin-2 fluctuations were studied in the seed $\text{AdS}_3$ massive type IIA solutions. \Cref{eq: ct_final}  is exactly the same as \cite[equation 37]{Speziali:2019uzn} as expected.
\section{Epilogue}\label{sec: epilogue}
In this work we have investigated some aspects of spin-2 fluctuations around a warped type IIB solution of the form $\text{AdS}_2 \times \text{S}^2 \times \text{CY}_2 \times \mathcal{I}_{\psi} \times \mathcal{I}_{\rho}$ of \cite{Lozano:2020txg}, which is related via T-duality to the $\text{AdS}_3 \times \text{S}^2 \times \text{CY}_2 \times \mathcal{I}_{\rho}$ background of \cite{Lozano:2019emq}. We have derived the equation that governs the dynamics of these spin-2 modes following the logic of \cite{Bachas:2011xa}. We have seen that they fall into two big classes, the universal and non-universal. The former are independent of the background supergravity data and are present in all the $d=1$ field theories that are associated with the backgrounds considered here.

As we have seen in the main body of our work, when the universal class of fluctuations saturates the bound on the mass, it is dual to operators with scaling dimension $\Delta = \ell+1$, where $\ell$ is the angular-momentum-charge on the $\text{S}^2$. This is the holographic manifestation of the $SU(2)_R$ symmetry of the dual field theory. 

The non universal solutions are more difficult to analyse as they depend on background data. In other words, one has to make specific choices for functions $u$, $\hat{h}_4$ and $h_8$ that fully determine the $\text{AdS}_2$ solutions. 

Finally, we have computed the central charge for the one-dimensional dual quiver field theory using the action for the spin-2 fluctuations, $h_{\mu \nu}$. 

As we have, already, seen these $\text{AdS}_2$ backgrounds share certain similarities in their structures with their higher-dimensional seed solutions. We have seen in detail the relations of the minimal universal class of modes to the corresponding ones in the $\text{AdS}_3$ solutions and how the central captures the correct behaviour of the two-dimensional seed conformal field theory. We have, also, discussed these similarities in some particular non-universal solutions. 

It would be interesting to take inspiration from the various $\text{AdS}_3$ studies and compute additional holographic observables to understand the $\text{AdS}_2 \times \text{S}^2 \times \text{CY}_2 \times \mathcal{I}_{\psi} \times \mathcal{I}_{\rho}$ more thoroughly. It would also be interesting to consider metric fluctuations in the $\text{T}^4$ part of the geometry. Since, none of the defining functions of the background depend on those directions these fluctuations would, also, satisfy a massless Klein-Gordon equation in ten dimensions, and would capture part of the spin-0 spectrum of the dual $\mathcal{N}=4$ quantum mechanical theory\footnote{we are grateful to Christoph Uhlemann for suggesting this point and related comments.}. It would be interesting to study these in our setup as well as in the seed $\text{AdS}_3$ and spot any similarities or differences. Finally, having already a map that takes us from $\text{AdS}_2$ to the $\text{AdS}_3$ solutions, it would be interesting to study the backgrounds that are described by $u = \texttt{constant}$ in $\text{AdS}_2$ and use the results in relation to the $\text{AdS}_3$ theories. 
\newpage
\section*{Acknowledgments}
I have greatly benefited from discussions and suggestions with Carlos Nunez. I am grateful to Nick J. Evans, Jose Manuel Penin and Stefano Speziali for related conversations. I would, also, like to thank Yolanda Lozano, Oleg Lunin, Carlos Nunez and Christoph Uhlemann for reading this manuscript and offering valuable comments. 
\newpage
\bibliographystyle{ssg}
\bibliography{spin2}
\end{document}